\documentclass[referee]{raa}            

\usepackage{graphicx,times}             
\usepackage{natbib}
\usepackage{amssymb,amsmath}
\bibpunct{(}{)}{;}{a}{}{,}

\usepackage[a4paper=true,dvipdfm=true,pagebackref=true]{hyperref}
\hypersetup{colorlinks = true, linkcolor = green, anchorcolor = red, citecolor = blue, filecolor = red, pagecolor = red, urlcolor = red}

\begin{document}

   \title{The Solar Upper Transition Region Imager (SUTRI) onboard the SATech-01 satellite
}

   \volnopage{Vol.0 (20xx) No.0, 000--000}      
   \setcounter{page}{1}          

   \author{Xianyong  Bai \inst{1,2} \and Hui Tian \inst{3,1,8*} \and Yuanyong Deng\inst{1,2} \and Zhanshan Wang \inst{4*} \and Jianfeng Yang \inst{5} \and Xiaofeng Zhang \inst{6} \and Yonghe Zhang \inst{6} \and  Runze Qi \inst{4} \and Nange Wang \inst{5} \and Yang Gao \inst{6} \and Jun Yu \inst{4} \and Chunling He \inst{4} \and Zhengxiang Shen \inst{4} \and Lun Shen \inst{5} \and Song Guo \inst{5} \and Zhenyong Hou \inst{3} \and Kaifan Ji \inst{7} \and Xingzi Bi \inst{6} \and Wei Duan \inst{1} \and Xiao Yang \inst{1} \and Jiaben Lin \inst{1} \and Ziyao Hu \inst{1} \and Qian Song \inst{1,2} \and Zihao Yang \inst{3} \and Yajie Chen \inst{3} \and Weidong Qiao \inst{5} \and Wei Ge \inst{5} \and Fu Li \inst{5} \and Lei Jin \inst{5} \and Jiawei He \inst{5} \and Xiaobo Chen \inst{5} \and Xiaocheng Zhu \inst{6} \and Junwang He \inst{6} \and Qi Shi \inst{6} \and Liu Liu \inst{6} \and Jinsong Li \inst{6} \and Dongxiao Xu \inst{6} \and Rui Liu \inst{6} \and Taijie Li \inst{6} \and Zhenggong Feng \inst{6} \and Yamin Wang \inst{6} \and Chengcheng Fan \inst{6} \and Shuo Liu \inst{6} \and Sifan Guo \inst{1,2} \and Zheng Sun \inst{3} \and Yuchuan Wu \inst{1,2} \and Haiyu Li \inst{3} \and Qi Yang \inst{2,3} \and Yuyang Ye \inst{1,2} \and Weichen Gu \inst{4} \and Jiali Wu \inst{4} \and Zhe Zhang \inst{4} \and Yue Yu \inst{4} \and Zeyi Ye \inst{4} \and Pengfeng Sheng \inst{4} \and Yifan Wang \inst{4} \and Wenbin Li \inst{4} \and Qiushi Huang \inst{4} \and Zhong Zhang \inst{4}
   }

   \institute{National Astronomical Observatories, Chinese Academy of Sciences, Beijing, 100101, China;\\
        \and
          School of Astronomy and Space Science, University of Chinese Academy of Sciences, Beijing 101408, China \\
          \and
          School of Earth and Space Sciences, Peking University, Beijing, 100871, People's Republic of China  {\it huitian@pku.edu.cn}\\
          \and
          	Tongji University, Institute of Precision Optical Engineering, School of Physics Science and Engineering, Key Laboratory of Advanced Micro-Structured Materials MOE, Shanghai, China  {\it wangzs@tongji.edu.cn}\\
          \and
            Xi'an Institute of Optics and Precision Mechanics, Chinese Academy of Sciences, Xi'an 710119, China\\
           \and 
           Innovation Academy for Microsatellites, Chinese Academy of Sciences, Shanghai 201304, China \\
           \and
           Yunnan Observatories, Chinese Academy of Sciences, Kunming, 650011, Yunnan, China \\
           \and
           Key Laboratory of Solar Activity and Space Weather, National Space Science Center, Chinese Academy of Sciences, Beijing 100190, China\\
\vs\no
   {\small Received~~20xx month day; accepted~~20xx~~month day}}

\abstract{The Solar Upper Transition Region Imager (SUTRI) onboard the Space Advanced Technology demonstration satellite (SATech-01), which was launched to a sun-synchronous orbit at a height of $\sim$500 km in July 2022, aims to test the on-orbit performance of our newly developed Sc/Si multi-layer reflecting mirror and the 2k$\times$2k EUV CMOS imaging camera and to take full-disk solar images at the Ne VII 46.5 nm spectral line with a filter width of $\sim$3 nm. SUTRI employs a Ritchey-Chrétien optical system with an aperture of 18 cm. The on-orbit observations show that SUTRI images have a field of view of $\sim$ 41.6’$\times$41.6’ and a moderate spatial resolution of $\sim$8” without an image stabilization system. The normal cadence of SUTRI images is 30 s and the solar observation time is about 16 hours each day because the earth eclipse time accounts for about 1/3 of SATech-01’s orbit period. Approximately 15 GB data is acquired each day and made available online after processing. SUTRI images are valuable as the Ne VII 46.5 nm line is formed at a temperature regime of $\sim$0.5 MK in the solar atmosphere, which has rarely been sampled by existing solar imagers. SUTRI observations will establish connections between structures in the lower solar atmosphere and corona, and advance our understanding of various types of solar activity such as flares, filament eruptions, coronal jets and coronal mass ejections.
}

   \authorrunning{X. Y. Bai, H. Tian, \& Y.Y. Deng, et al. }            
   \titlerunning{SUTRI }  

   \maketitle

%
%
\section{Introduction}

The solar upper atmosphere, the corona and transition region (TR), plays a critical role in channeling mass and energy from our Sun to the interplanetary space. The million-kelvin solar corona can be observed on the ground during total solar eclipses \citep[e.g., ][]{Quzq2013,Chenyj2018} or with coronagraphs \citep[e.g., ][]{Yangzh2020a,Yangzh2020b,Zhangxf2022}. However, these observations can only reveal coronal structures and dynamics above the limb of the solar disk. The corona and TR on the front side of the Sun cannot be observed at the optical and infrared wavelengths. Being much hotter than the Sun’s lower atmosphere, the photosphere and chromosphere ($\le$0.02 MK), the corona and TR are known to produce many strong emission lines at the extreme ultraviolet (EUV) wavelengths \citep{Delzanna2018}. These lines are usually optically thin and the lower atmosphere does not emit these lines. Thus, observations with these lines could reveal structures and dynamics of the upper atmosphere both on the front side and in off-limb regions of the Sun.

EUV observations of the Sun can be dated back to the 1960s, when pinhole cameras with gratings carried by the Skylark rockets recorded solar images \citep{Burton1969}. In the past half century, dozens of EUV imagers and spectrographs have been sent to space to observe the Sun. The spatial resolution and effective area have been increasing over the years. These EUV telescopes were built with either a grazing-incidence system or normal-incidence system. The latter normally leads to better spatial resolution. However, normal-incidence systems are subject to lower throughput due to the low reflectivity of most materials for EUV radiation \citep[e.g., ][]{Wilhelm2004,bai2023}. To increase the reflectivity, the multi-layer technique has often been adopted in the past three decades. The Extreme-ultraviolet Imaging Telescope \citep[EIT,][]{Delaboundini1995} onboard the Solar and Heliospheric Observatory (SOHO), which was launched in December 1995, marked a milestone for EUV observations of the Sun as it took full-disk solar images with four narrow-band filters (He II 30.4 nm, Fe IX/X 17.1 nm, Fe XII 19.5 nm, Fe XV 28.4 nm) in a routine way. The primary mirror is coated with Mo-Si multilayers. The spatial resolution of the obtained imager is about 6”. The same passbands were adopted later by the Extreme UltraViolet Imager \citep[EUVI,][]{Wuelser2004} onboard the twin spacecraft of the Solar Terrestrial Relations Observatory (STEREO), with a spatial resolution of ~4”. Higher-resolution (~1.5”) EUV images with a field of view (FOV) of 8.5’x8.5’ were taken by the Transition Region and Coronal Explorer \citep[TRACE,][]{Handy1999}. The three EUV narrow passbands are centered at 17.1 nm, 19.5 nm and 28.4 nm with a bandwidth of 0.64 nm, 0.65 nm and 1.07 nm, respectively. Since 2010, the Atmospheric Imaging Assembly \citep[AIA,][]{Lemen2012} onboard the Solar Dynamics Observatory \citep[SDO,][]{Pesnell2012} has been taking full-disk images in ten passbands with a similar resolution. Seven of these ten passbands fall into EUV wavelengths, i.e., 17.1 nm, 19.3 nm, 21.1 nm, 33.5 nm, 9.4 nm, 13.1 nm and 30.4 nm. AIA normally takes images in each of these passbands at a high cadence of 12 s. EUV images with a sub-arcsec resolution were first taken by two rocket flights of the High Resolution Coronal Imager \citep[Hi-C][]{Kobayashi2014, Rachmeler2019} in 2012 and 2018. The two flights obtained coronal images at a spatial resolution of 0.2-0.3” with the 19.3 nm and 17.2 nm filters, respectively. More recently, as the Solar Orbiter \citep[SO,][]{Muller2020} gets close to the Sun, one of its payload the Extreme Ultraviolet Imager \citep[EUI,][]{Rochus2020} also takes solar EUV images with the highest spatial resolution close to that of Hi-C in the Fe X 17.4 nm passband. 

The EUV imagers mentioned above are generally targeting at wavelengths shorter than 35 nm. Except for the He II 30.4 nm passband that is mostly sensitive to plasma cooler than 0.1 MK, these previously adopted EUV passbands are all sensitive to emission from the coronal plasma hotter than 0.8 MK. Obviously, there is a temperature gap between the “hot” and “cool” bands. This is exactly the upper TR, the temperature regime of 0.1-0.8 MK. As discussed in \citet{Tian2017} and \citet{Tian2021}, this temperature regime is not only important for our understanding of the mass and energy transport mechanisms in the solar atmosphere, but also critical for diagnosing various physical processes involved in solar eruptions. So routine narrow-band imaging of this temperature regime is highly desired. 

One of the strongest emission lines formed in this temperature regime is the Ne VII 46.5 nm line, which has a formation temperature of ~0.5 MK under ionization equilibrium \citep[e.g.,][]{Tian2017,Chenyj2022}. However, this line has not been intensively used in solar observations. The last time that full-disk solar images of this line were obtained was about half century ago. The slitless EUV grating spectrograph S082A onboard Skylab obtained some Ne VII 46.5 nm images in 1973 \citep{Tousey1973}. However, due to the intrinsic shortcoming of slitless spectrographs, these images suffered from a mixture of spectral and spatial information. The grazing-incidence spectrometer of Coronal Diagnostic Spectrometer \citep[CDS,][]{Harrison1995} onboard SOHO had the capability of obtaining solar spectra in the wavelength range of 15.1-78.5 nm, covering the Ne VII 46.5 nm line. However, the dimensions of the slits are 2”x2”, 4”x4” and 8”x50”, making high-resolution or large-FOV images of Ne VII 46.5 nm impossible to obtain. With the aim of producing simultaneous spatial-spectral images of Ne VII 46.5 nm, the rocket flight of the Multi-Order Extreme Ultraviolet Spectrograph (MOSES) in August 2015 recorded solar images for three spectral orders (0, +1, -1) \citep{Fox2011,Smart2016}. However, rocket observations only lasted for a period of about ten minutes and the disambiguation of the spatial and spectral information is not an easy task. 

Based on a mission concept study and an advanced research of space science missions supported by the strategic priority program on space science, Chinese Academy of Sciences (CAS), we proposed that narrow-band imaging and high-resolution spectroscopic observations of the upper TR with spectral lines such as Ne VII 46.5 nm should be the top priority of future TR observations \citep{Tian2017,bai2023}. In December 2019 CAS announced an opportunity of carrying instruments on the first satellite platform of the Space Advanced Technology demonstration satellite series (SATech-01). We decided to make use of this opportunity and eventually built a narrow-band Ne VII 46.5 nm imager, the Solar Upper Transition Region Imager (SUTRI). SATech-01 was successfully launched to a sun-synchronous orbit at a height of ~500 km in July 2022 by the Zhongke-1A (ZK-1A or Lijian 1) rocket at the Jiuquan Satellite Launch Center. The aperture door of SUTRI was opened on August 31. Since September 4, 2022, SUTRI has been routinely taking narrow-band images of the solar upper TR during the roughly two thirds of each orbit period ($\sim$60/96 min).

In this paper we present an overview of the scientific objectives, instrument design, ground test results, on-orbit operation, calibration and data reduction of SUTRI. The early results from SUTRI are also presented.


\section{Science with the Ne VII 46.5 nm line}
As mentioned above, the Ne VII 46.5 nm line is formed at a temperature regime of $\sim$0.5 MK in the solar atmosphere, which has rarely been sampled before. Higher-temperature plasma in the solar atmosphere has been routinely observed with SDO/AIA and many other telescopes. Many ground-based and space-born telescopes have also been monitoring the lower solar atmosphere including the photosphere and chromosphere \citep[e.g.,][]{Liu2014, Yan2020, Fang2013}. The lower TR has also been routinely observed with the AIA 30.4 nm filter and the Interface Region Imaging Spectrograph mission \citep[IRIS,][]{de2014}. With a unique filter centered at the Ne VII 46.5 nm line, SUTRI fills in the key temperature gap of current solar observations. Thus, SUTRI images will provide key information that is highly complementary to that obtained by other solar telescopes such as SDO/AIA and IRIS. Although SUTRI is a low-cost experiment, we still anticipate that its observations will advance our understanding of the mass and energy transport processes in the solar atmosphere.

\begin{figure}
    \centerline{\includegraphics[width=0.75\textwidth, angle=0]{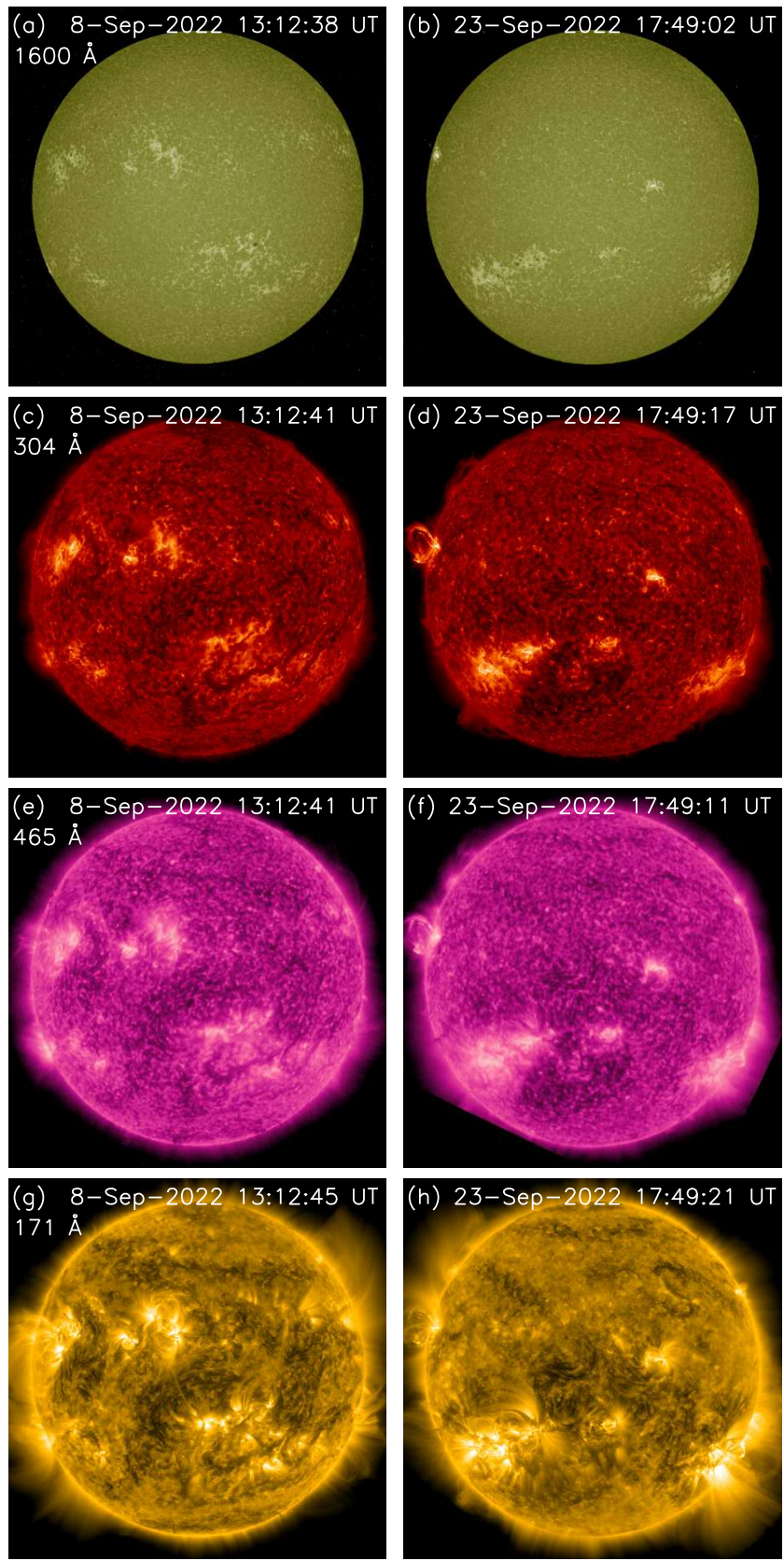}} \caption{ \textbf{46.5 nm, 160 nm, 30.4 nm and 17.1 nm images taken by SUTRI and AIA on 8 Sep. and 23 Sep., 2022.}  }
    \label{fig.1}
\end{figure}
 
\subsection{Establish the missing link between the lower solar atmosphere and corona}
The mass that fills in the corona and the energy that powers the corona both come from the lower solar atmosphere. So the coupling between the lower atmosphere and corona is of vital importance for our understanding of coronal heating and origin of the solar wind. Previous imaging observations have clearly revealed that structures cooler than 0.1 MK are largely different from those hotter than 0.8 MK in not only active regions (ARs) but also quiet-Sun regions \citep{Tian2017}. Obviously, the missing link between the two lies in the upper TR. The Ne VII 46.5 nm images obtained by SUTRI clearly reveal solar structures such as the quiet-Sun network, AR loops and coronal holes (seen from Figure\ref{fig.1}). The typical size of network lanes appear to be larger than that observed in the 160 nm passband of SDO/AIA, which could be explained by the expansion of magnetic funnels with height \citep{Dowdy1986,Tian2008}. We also found that the observed AR loop systems appear to be in the transition state between the loops in AIA 17.1 nm images and their counterparts in AIA 30.4 nm. Coronal holes are also clearly observed, but their contrast with respect to the surrounding quiet-Sun region appears to be smaller than that in AIA 17.1 nm images. In a word, SUTRI observes solar structures that are in the transition state from the lower TR to the corona, confirming that the Ne VII 46.5 nm line is formed in the upper TR.

The upper TR is different from the lower TR and corona in not only the emission structures but also the flow patterns. Previous spectroscopic observations have revealed a predominance of redshifts in most TR spectral lines. There is a clear trend that the average Doppler shift increases from almost zero at a temperature of $\sim$0.02 MK to a red shift of 10 km/s at $\sim$0.2 MK in the quiet Sun. As the temperature continues to increase, the red shift starts to decrease and finally turns into blue shift at temperatures typical of the lower corona \citep[e.g.,][]{Peter1999,Teriaca1999, Xia2004, Wang2013}. The temperature at which the Doppler shift changes its sign is around 0.5 MK, which is exactly the formation temperature of the Ne VII 46.5 nm line. This sign change has been recently reproduced through a three-dimensional radiative hydrodynamic simulation \citep{Cyj2022}. Since these systematic flows likely result from various heating and cooling processes in the solar atmosphere \citep{Hansteen2010,Peter2006}, the Ne VII 46.5 nm line carries important information to improve our understanding of the coronal heating mechanisms. Moreover, the blue shifts of spectral lines formed at higher temperatures have been suggested to be signatures of the nascent solar wind \citep[e.g.,][]{Hassler1999, Xia2003, Tu2005, He2008, Tian2010}. If this is the case, the Ne VII 46.5 nm line could be used to directly probe the source regions of the solar wind, since the layer of 0.5 MK marks the boundary between downflows and outflows in the solar atmosphere. Although SUTRI could not conduct spectroscopic observations, the direct imaging of plasma flows with a temperature of 0.5 MK could still reveal valuable information of physical processes involved in heating of the corona and generation of the solar wind.

The prevalent small-scale intermittent chromospheric jets, or spicules, have also been suggested to supply hot materials to the corona \citep{DePontieu2011,Samanta2019}. However, the coronal passbands of AIA used in these studies contain emission from not only hot coronal lines but also cool TR lines. Because of this complexity, the claim of spicule heating to million kelvin has been challenged \citep[e.g.,][]{Madjarska2011, Klimchuk2012}. Nevertheless, the imaging and spectroscopic observations of IRIS have already demonstrated that some spicules are heated to at least $\sim$0.1 MK \citep{Tian2014,Pereira2014}. The Ne VII 46.5 nm line is the only strong line in the passband of 46.5$\pm$1.5 nm. With a narrow-band 46.5 nm imager, in principle one can track the continued heating of spicules to a temperature of $\sim$0.5 MK, if such heating does exist. Moreover, previous imaging observations with SDO/AIA and spectroscopic observations with the EUV imaging spectrometer \citep[EIS,][]{Culhane2007} onboard Hionde have implied the following scenario of chromosphere-corona mass cycle or circulation \citep{Marsch2008,McIntosh2012}: chromospheric spicules are intermittently and rapidly heated to million kelvin, followed by slow but efficient radiative cooling of the draining material. With a formation temperature of $\sim$0.5 MK, the Ne VII 46.5 nm line is an excellent one for the diagnostics of the cooling downflows. From SUTRI observations, such downflows have already been identified and appear to be common along loop legs at AR boundaries. Future analysis of these downflows will likely improve our understanding of this mass cycling process.

\subsection{Investigate the dynamics and energetics of solar eruptions}

SUTRI observations have already demonstrated that the Ne VII 46.5 nm line is an excellent one for monitoring solar flares, coronal mass ejections (CMEs), prominences/filaments and their eruptions. For instance, filaments are clearly observed as absorption features in the 46.5 nm images, based on which the mass of filaments might be estimated \citep[e.g.,][]{Jenkins2018}. SUTRI is also expected to capture dynamic events such as coronal condensation \citep[e.g.,][]{Li2018,Chenhc2022}, twisting/untwisting motions of filaments \citep[e.g.,][]{Xue2016}, solar giant tornadoes \citep[e.g., ][]{Su2012,Yang2018}, interaction between falling prominence material with the lower atmosphere \citep[e.g.,][]{Su2015,Li2017}, and magnetic reconnection between filaments and other coronal structures \citep[e.g.,][]{Li2016}.

Past and current EUV imagers mostly use spectral lines formed at temperatures typical of the chromosphere, lower TR and corona to monitor solar eruptions. With these observations, we have a relatively good understanding of the energetics of plasma hotter than $\sim$0.8 MK and cooler than $\sim$0.1 MK at eruption sites or within eruptive structures. SUTRI observations fill in the temperature gap of 0.1-0.8 MK, which will undoubtedly lead to a more complete and accurate understanding of the dynamics and energetics of solar eruptions when coordinated with observations from other telescopes such as SDO/AIA.

For instance, observations of flare regions with other telescopes reveal either flare ribbons or loops \citep[e.g.,][]{Tian2016}. Due to the lack of observations of the 0.1-0.8 MK plasma, intermediate structures between the ribbons and loops are generally not sampled. SUTRI observations will likely reveal intermediate structures, and link the heating processes at flare ribbons and loop tops. 

SUTRI could also improve our understanding of the temperature structures at flare sites and within CMEs. In the past decade, observations from the six Fe-dominated passbands of SDO/AIA have been frequently used for the inversion of differential emission measure (DEM) \citep[e.g.,][]{Weber2004,Hannah2012,Cheng2012,Cheung2015,Su2018}. However, these DEM curves are not well constrained at temperatures lower than 0.8 MK \citep[e.g.,][]{Samanta2021}, mainly due to the weak response of these passbands at such low temperatures. If supplemented with simultaneous observations from SUTRI, the low-temperature part should be better constrained. As a result, the energetics of the observed eruptive events can be better understood. 

Since SUTRI is sampling the layer between the chromosphere and corona, its observations may provide important clues for the connection between EUV waves \citep[e.g.,][]{Thompson1999,Chen2002,Zheng2022} and Moreton-Ramsey waves triggered by solar eruptions \citep[e.g.,][]{Moreton1960}. Although it has been suggested that Moreton-Ramsey waves are chromospheric imprints of coronal EUV waves, simultaneous observations of the two are very rare, making it difficult to prove this idea \citep[e.g.,][]{Shen2012,Hou2022}. In addition, it is not straightforward to establish their connection as the waves are normally observed at largely different heights of the solar atmosphere. Now with continuous seeing-free chromospheric observations from the Advanced Space-based Solar Observatory \citep[ASO-S or Kuafu-1,][]{Gan2019} and the Chinese H$\alpha$ Solar Explorer \citep[CHASE,][]{Li2022}, we expect to observe many more events of Moreton-Ramsey wave. SUTRI observations will provide key information to establish the link between these Morrton-Ramsey waves and the EUV waves observed with coronal imagers such as SDO/AIA, STEREO/EUVI and the X-ray Extreme Ultraviolet Imager \citep{Chen2022,song2022}  onboard the Fengyun-3E meteorological satellite. These observations will undoubtedly improve our understanding of the nature of EUV waves and the relationship between the two wave-like phenomena. 

\section{Overview of the instrument}

One of the purposes of SUTRI is to test the on-orbit performance of our newly developed Sc/Si multilayer mirror and the back-illuminated CMOS imaging camera on the SATech-01 satellite.  The Sc/Si multilayer mirror working at 46.5 nm has never been used by past and existing EUV imagers, and the CMOS imaging camera is used for the first time in EUV astronomical observations of China. Meanwhile, SUTRI is designed to take full-disk images of the solar atmosphere at a temperature of $\sim$0.5 MK to realize the above-mentioned scientific objectives if the multilayer mirrors and CMOS camera work well on orbit. We also try to develop some on-orbit calibration methods to improve the science data quality as China has conducted very few EUV astronomical observations in the past. All of these attempts will be extremely important for future solar EUV telescopes or EUV space missions. SATech-01 satellite has 16 payloads including several remote sensing payloads to observe the earth at the visible and infrared wavelengths, and to observe the universe at soft x-ray \citep{Zhangc2022}. Each orbit period (96 min) of SATech-01 has about two thirds of time (60 min) for solar observations. The remaining time is arranged for the other payloads’ observations. The average solar observation time is about 16 hours per day and the data rate for SUTRI is about 15 GB per day. The pointing accuracy and the stability of the satellite is about 0.005$^\circ$ and 0.0005 $^\circ/s$, respectively. Based on these, the main instrument characteristics of SUTRI are determined and listed in Table 1. Due to the very limited financial support, manpower, time and the lack of special calibration equipment at 46.5 nm, SUTRI does not have an image stabilization system and many ground tests and improvement of instrument design could not be done.

\begin{figure}
    \includegraphics[width=\textwidth, angle=0]{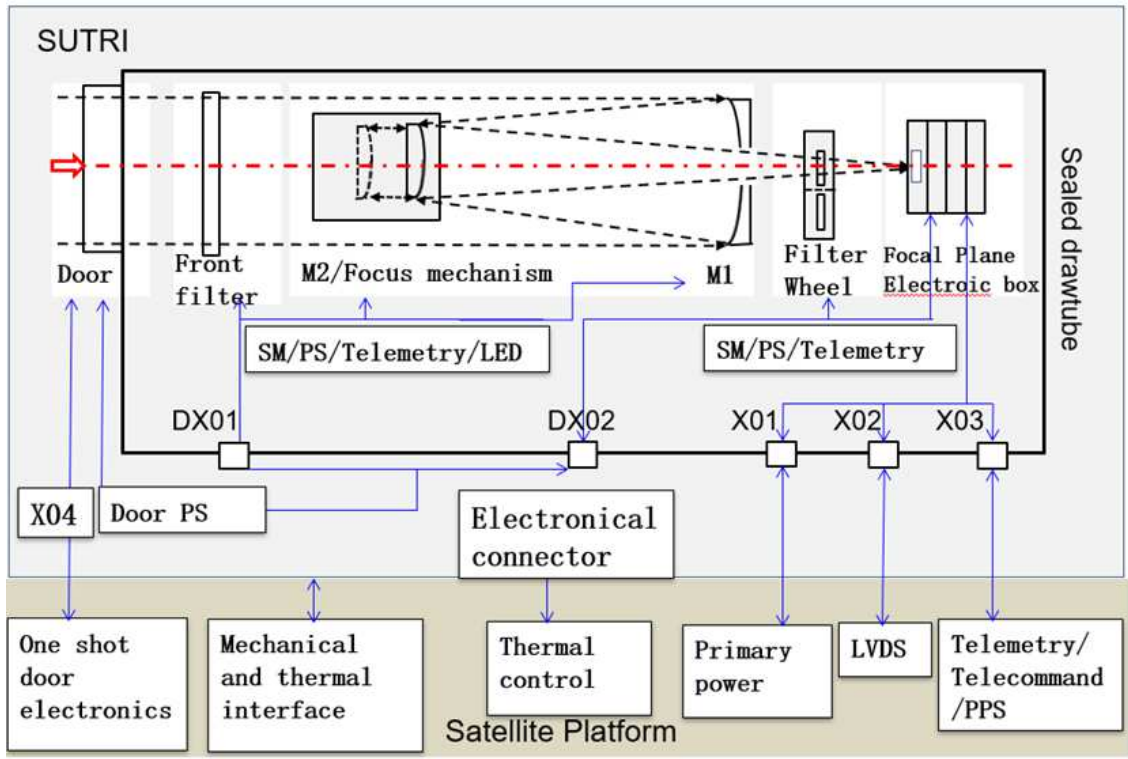} \caption{Block diagram of SUTRI. M1 and M2 indicates the primary and secondary reflecting mirrors, respectively. SM represents step motor while PS represents position switch. PPS is the pulse per second signal used for determining the time. }
    \label{fig.2}
\end{figure}

From the block diagram shown in Figure \ref{fig.2}, we can see that SUTRI consists of the following parts: 

\begin{itemize}
\item \textbf{It has a Ritchey-Chrétien optical system to image the Sun.} 
\item The front and rear filters are used to suppress the spectral contamination at other wavelengths, especially the strong visible and infrared solar radiation.
\item \textbf{An electronics box including the CMOS imaging sensor, the circuit boards and flight software are equipped to configure the observing parameters and to drive the motion mechanism.}
\item All of the mechanical structures supporting the two mirrors, the front filter, the rear filter wheel, the focus mechanism, the focal plane CMOS camera and its electronics box are arranged in a sealed drawtube with a sealed one-shot aperture door to avoid the degradation effects of contamination at EUV wavelengths, which is also employed in the previous EUV solar telescopes \citep{Lemen2012,de2014,Handy1999,Wuelser2004}. 
\item SUTRI also has a thermal control system to make sure that every component works at an appropriate temperature. 
\end{itemize}

As SUTRI is a UV instrument, stringent control of contamination is needed and we try our best to reduce its affect during the design, building and test phase. Moreover, an LED is added to monitor the performance of the detector and the position of the focus motion mechanism.

\begin{table}
    \begin{center}
        \caption[]{ Main instrument characteristics of SUTRI.}\label{tab1}
        \begin{tabular}{clclcl}
            \noalign{\smallskip}\hline
            Parameter &  Value/Mode \\
            \noalign{\smallskip}\hline
            Entrance aperture diameter & 18 cm  \\
            Field of view & $\sim$ 41.6’$\times$41.6’  \\
            Spatial resolution & $\sim$8”  ($\sim$1.22”/pixel) \\
            Working line & Ne VII 46.5 nm \\
            Formation temperature &	0.5 MK \\
            Filter width (FWHM)&	$\sim$3nm@46.5nm \\
            CMOS detector &	2k $\times$ 2k \\
            Photon numbers (in theory) &	Active Region: ~700 photons/pixel/s \\
                         &	 Quiet Sun: ~135 photons/pixel/s \\
            Observing mode &	Routine mode \\
             & Dark calibration mode \\ 
            & Flat field calibration mode \\
            & Cleaning mode \\
             Cadence &	Routine mode: 30s \\
                        Data rate &	15 GB/day \\
            Weight &	$\sim$30 Kg \\
            Power &	$\sim$30W (without thermal control system) \\
            
            \noalign{\smallskip}\hline
        \end{tabular}
    \end{center}
\end{table}	

\subsection{Optical and Mechanical structure}

SUTRI chooses the Ne VII 46.5 nm line to map the transition region at $\sim$ 0.5 MK. Similar to SOHO/EIT, TRACE, STEREO/EUVI and SDO/AIA, SUTRI adopts a normal-incidence optical system. The Ritchey-Chrétien system with two hyperboloid mirrors is used. The effective focal length is 1.09 m according to the 41.6$'\ \times$ 41.6$'$ FOV and the 6.5 $\mu m \times$ 6.5 $\mu m$ pixel size of the detector. To collect more photons, SUTRI's aperture diameter is 18 cm. Super-smooth primary and secondary mirrors are fabricated with a surface roughness better than \textbf{0.2 nm rms }to reduce the stray light and to enhance the reflectivity of multi-layer mirror at EUV wavelengths. We also add baffles on the supporting structures of the two mirrors to suppress stray light outside of SUTRI's FOV.

Both mirrors were coated by the Sc/Si multilayers to enhance the reflectivity at 46.5 nm. The Sc/Si multilayer sample with a period of 23.2 nm and a thickness ratio (the Sc layer thickness divided by the mutlilayer period) of 0.65 is firstly fabricated. The test result on the BESSY-II facility \citep{Sokolov2014} shows that a bandwidth (FWHM) of 3.68 nm is achieved with the peak wavelength at 45.5 nm \citep{Wu2022}. The reflectivity of the flight mirror is measured by the EUV reflectometer using line spectra at 46.1 nm generated by a radio frequency induced gas-discharge lamp at Tongji University \citep{Yu2022} at the incidence angle of 5°. The multilayer structure is further optimized to achieve higher reflectivity at 46.1 nm. The period and thickness ratio of the flight mirror are 24.55 nm and 0.68, respectively. The corresponding theoretical reflectivity is about 30.7$\%$ with a bandwidth of 3.7 nm.  
\begin{figure}
    \includegraphics[width=\textwidth, angle=0]{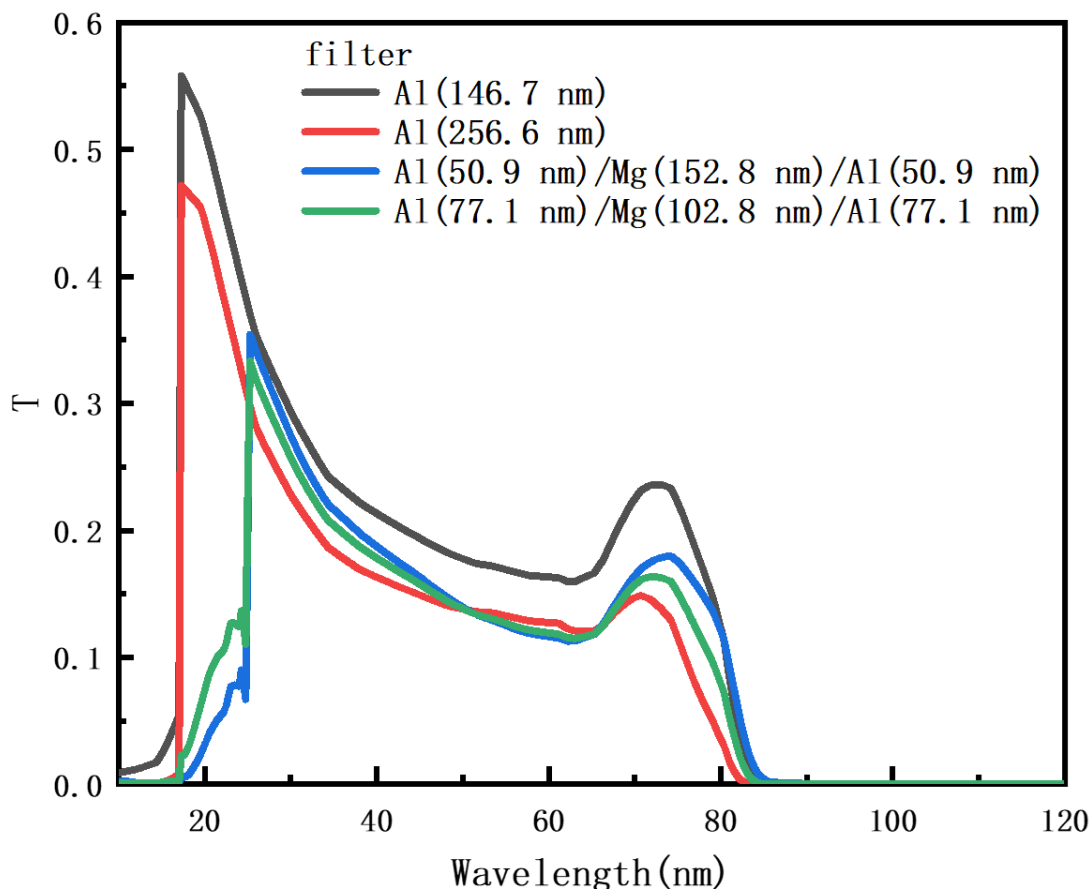} \caption{Theoretical transmission of the four Al and Al/Mg/Al rear thin film filters used by SUTRI. }
    \label{fig.3}
\end{figure}
\begin{figure}
    \includegraphics[width=\textwidth, angle=0]{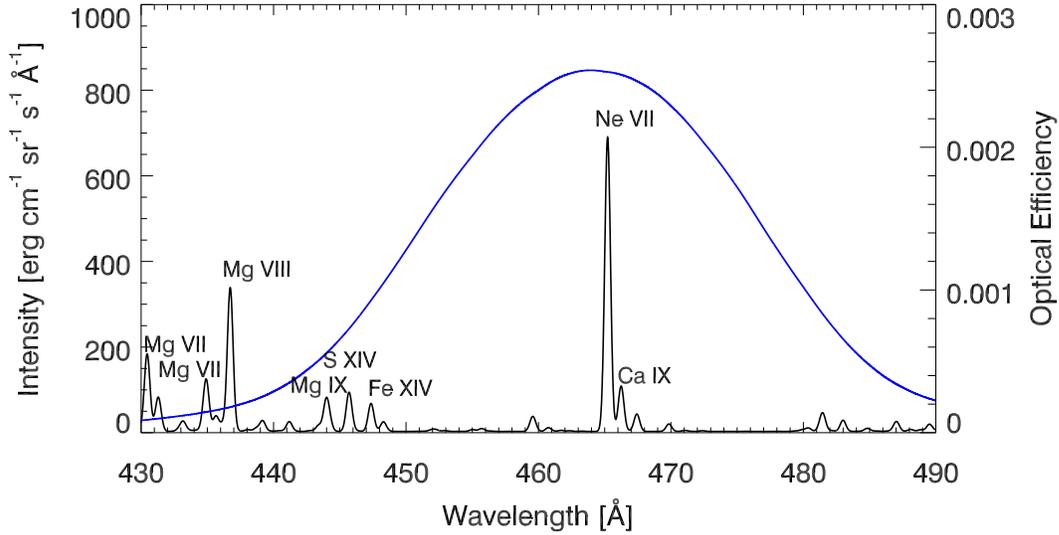} \caption{Synthetic solar spectrum near the Ne VII 46.5 nm line for an active region model \textbf{overplotted with the theoretical efficiency curve of SUTRI's optics including the two multilayer coating mirrors, the front and rear filters.}}
    \label{fig.4}
\end{figure}

The Sc/Si multilayer also has reflectivity at near ultraviolet, visible and infrared wavelength. In order to remove the strong solar visible and infrared wavelength, a high-purity Al front filter located near the door with the thickness of 200 nm is used by SUTRI. The size of the Al front filter is almost the same as the 18 cm aperture. Small rear filters are also adopted to suppress the out-of-band stray light caused by the inevitable pinholes in the front filter. The position of the rear filter is located in front of the detector. The reflectivity of Sc/Si multilayer mirror has a second peak near 21 nm, which is removed by the Al/Mg/Al filter as the absorption edge of Mg is about 26 nm. We installed two Al filters with the thickness of 150 nm and 200 nm and two Al/Mg/Al filters with the thickness of 50 /150/50 nm and 77/102/77 nm.  All thin film filters are supported by nickel mesh provided by LUXEL corporation. The theoretical transmission curves of four rear filters are presented in Figure \ref{fig.3}. The transmission of the front and rear filters is about 0.2 (ranging from 0.15 to 0.22) at 46.5 nm. \textbf{Only the Al/Mg/Al rear filter with the thickness of 77/102/77 nm is used in SUTRI's routine observation. The theoretical efficiency of SUTRI's optics considering the combination of the filters and multilayer mirrors is about 0.0025 at 46.5 nm according the following Formula:}
\begin{equation}\label{eq1}
    Efficiency =T_{front\  filter} \times R_{Primary\  mirror} \times R_{secondary\  mirror} \times T_{rear\  filter} \\
    \approx 0.18 \times 0.3 \times 0.3 \times 0.16=0.0025.
\end{equation}

There is no ground calibration facility in China that can accurately measure the reflectivity and transmission curve of the thin film filter and multilayer mirror at 46.5 nm. Consequently, we did not obtain the measured efficiency curves of SUTRI during the groud calibration. With the theoretical efficiency \textbf{of SUTRI's optics shown in Figure \ref{fig.4}} and assuming that the quantum efficiency of CMOS detector is 0.2 (see Section 3.3.3 for details), the photon numbers are estimated for both active and quiet sun models and listed in Table \ref{tab1}. \textbf{We also plot the synthetic solar spectrum near the 46.5 nm for an active region model with the Chianti database \citep{DelZanna2021,Dere1997}. It clearly shows that the Ne VII 46.5 nm line dominates the wavelength integrated radiation within the bandwidth of the spectral response curve. The bandwidth is about 3 nm. }   

\begin{figure}
    \includegraphics[width=\textwidth, angle=0]{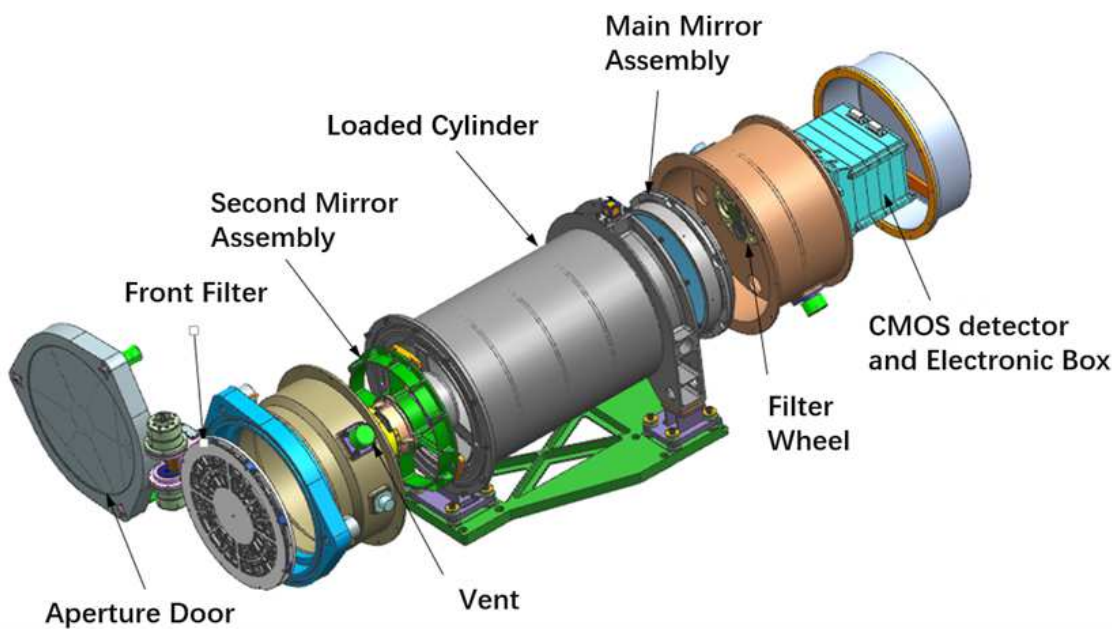} \caption{Exploded diagram of SUTRI's optical and mechanical structure.}
    \label{fig.5}
\end{figure}

\begin{figure}
    \includegraphics[width=\textwidth, angle=0]{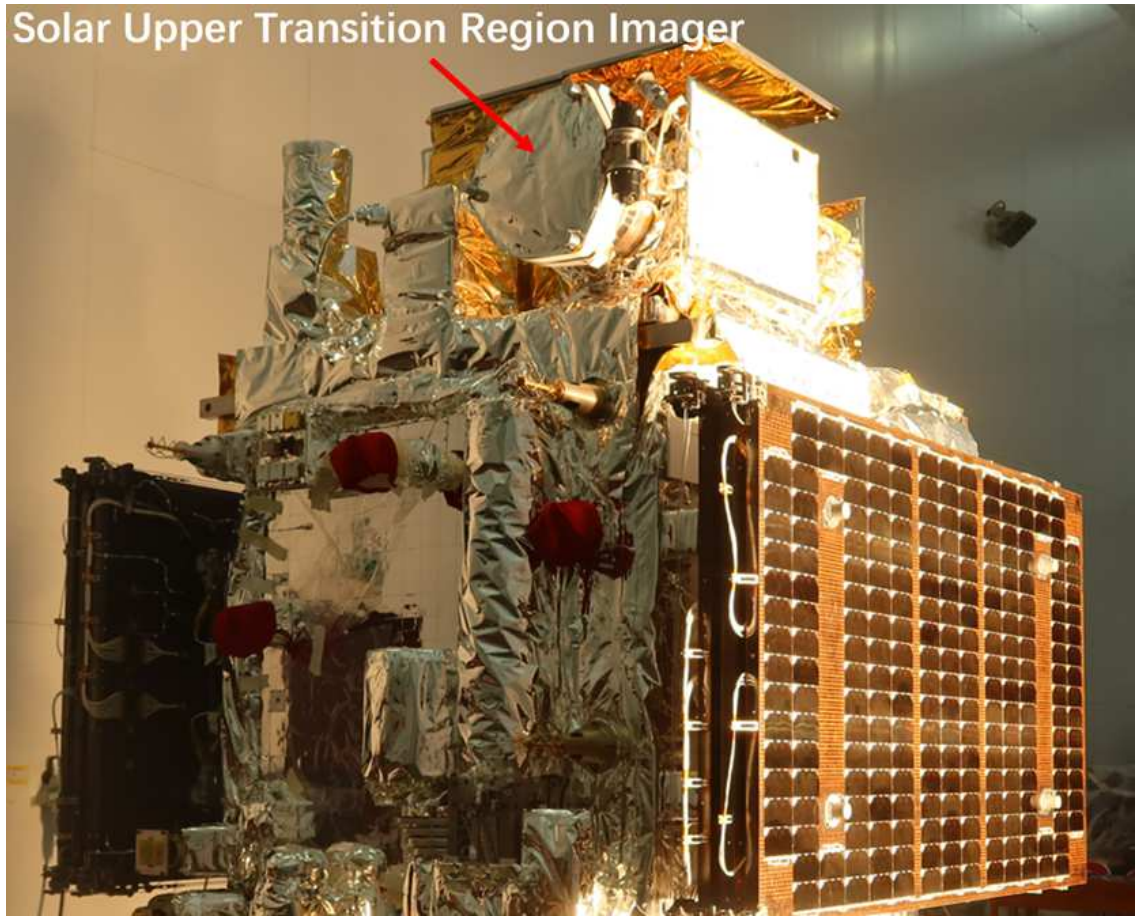} \caption{SUTRI installed on the SATech-01 Satellite. }
    \label{fig.6}
\end{figure}

Figure \ref{fig.5} presents the exploded diagram of SUTRI's optical and mechanical structure. SUTRI has an aperture door, a front filter assembly, a front tube, a loaded cylinder, a back tube and a rear cover from front to back. The primary and secondary mirror assembly are integrated on the loaded cylinder. SUTRI connects with the SATech-01 satellite by three supporting legs located at the bottom face of the loaded cylinder. The electronic box is integrated with SUTRI's back tube and is softly connected with the rear cover. SUTRI has three motion mechanisms. The one-shot door mechanism driven by the reusable shape memory alloy. The focus can be adjusted by moving the position of the second mirror with a stepper motor to compensate for the possible changes of on-orbit focus. Four rear filters are selected by the filter wheel mechanism. The length, width and height of SUTRI after integration are about 85 cm, 38 cm and 38 cm, respectively. A picture of SUTRI installed on the satellite is shown in Figure \ref{fig.6}.  

\begin{figure}
    \includegraphics[width=\textwidth, angle=0]{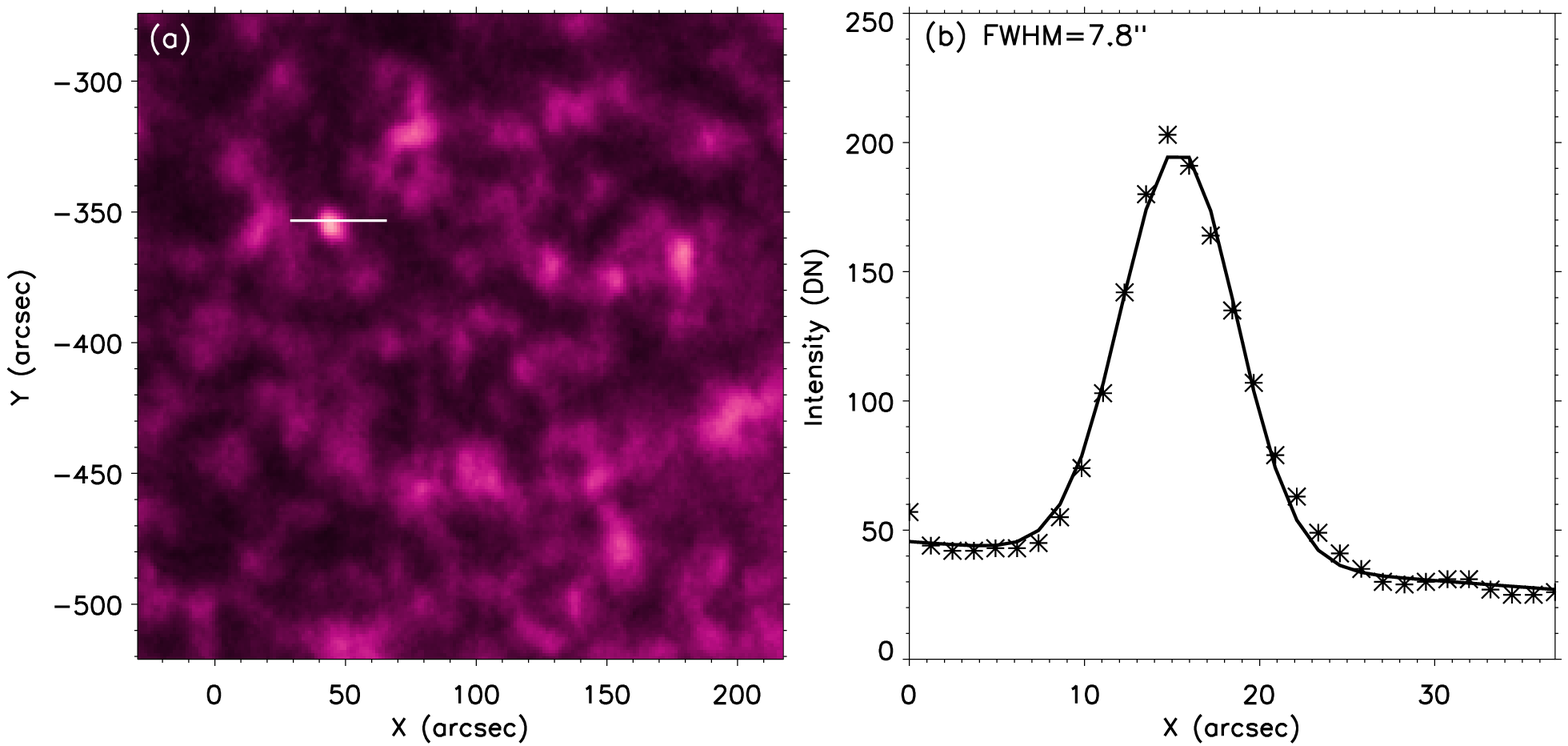} \caption{Panel (a): A quiet-sun region observed by SUTRI. Panel (b): Intensity distribution along the horizontal slice across the bright point in panel a. A Gaussian fitting (solid line) is used to estimate the spatial resolution.}
    \label{fig.7}
\end{figure}

The spatial resolution of SUTRI is about 6" during the ground calibration with the collimating light from a Zygo interferometer in the air. An image of the quiet sun region taken by SUTRI on October 10, 2022 is shown in Figure \ref{fig.7}. The on-orbit spatial resolution is evalued with the isolated bright point. We selected a slice on the bright point (the horizontal line in Figure \ref{fig.7}a), conducted a single gaussian fitting and calculated the full width at half maximum. The corresponding result is 7.8". Thus, we concluded that SUTRI images have a moderate resolution of $\sim$8”. From the apparent diameter of the solar disk, we can estimate the FOV of SUTRI, which is \textbf{$\sim$41.6’$\times$41.6’} with a pixel resolution of $\sim$1.22”/pixel. More details about SUTRI's telescope design and optical structure are given by Wang et al., 2023.

\subsection{Thermal}

SUTRI employs both active and passive thermal control methods to meet its working temperature requirements, which are listed in Table 2. The design of the thermal control is verified by the ground thermal balance test. We also show the on-orbit results in Table 2. Details about the thermal control system of SUTRI can be found in Gao et al., 2023. 

\begin{figure}
    \includegraphics[width=\textwidth, angle=0]{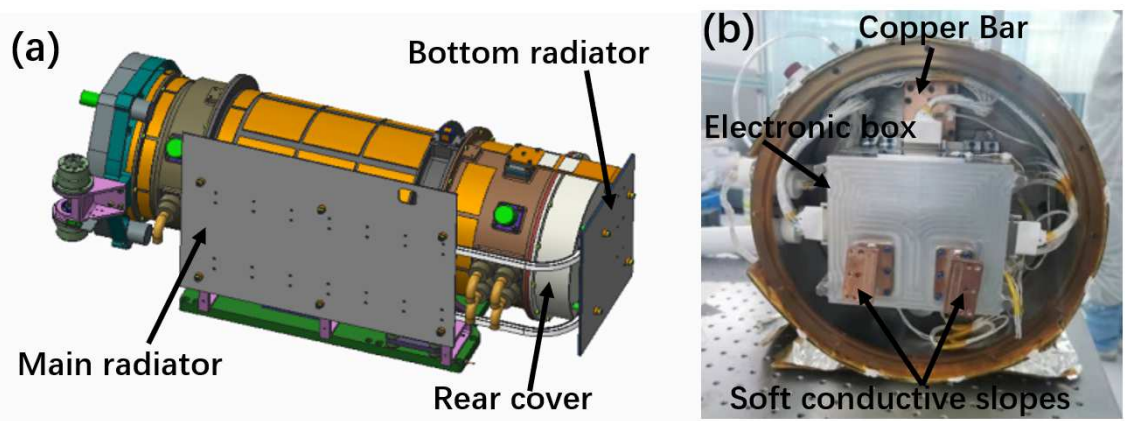} \caption{Panel (a): Heaters (orange color) on the main structure. Panel (b): Soft conductive slopes on rear surface the electronic control box. }
    \label{fig.8}
\end{figure}

To avoid contamination, all of the heaters and most of the thermistors are arranged on the outer side of SUTRI’s main structure. Internal optical system including the primary mirror, secondary mirror, the loaded cylinder, the front tube, the focus adjusting mechanism and the filter wheel, is indirectly controlled in the range of 20$\pm$2$ ^\circ$C by the heaters on the main structure (see Figure \ref{fig.8}a). The rear cover at the end of the main structure is innovatively set as a cold plate, which connects the main radiator with two heat pipes and can collect parts of the on-orbit contamination released from SUTRI’s internal structures. The on-orbit working temperature of the rear cover is between -30 and -15℃. The dissipation of the heat flux from the electronic control box is conducted to the rear cover by two soft conductive slopes, as shown in Figure \ref{fig.8}b.

\begin{table}
    \begin{center}
        \caption[]{ Requirement and on-orbit performance of SUTRI’s thermal control system.}\label{tab2}
        \begin{tabular}{clclcl}
            \noalign{\smallskip}\hline
            Parameter &	Requirement	& On-orbit result \\
            \noalign{\smallskip}\hline
            Front filter &	0$\sim$50$ ^\circ$C &	10$\sim$40$ ^\circ$C \\
            loaded cylinder	& 20$\pm$2$ ^\circ$C &	21$\pm$1$ ^\circ$C \\
            Primary mirror & 20$\pm$2$ ^\circ$C & 19$\pm$1$ ^\circ$C \\
            Secondary mirror & 20$\pm$2$ ^\circ$C & 21$\pm$1$ ^\circ$C \\
            Detector copper bar & Cleaning mode: $\sim$20$ ^\circ$C &
            	Cleaning mode: 20$\pm$1$ ^\circ$C \\
            & Working mode: -15 $\sim$ 0$ ^\circ$C &Working mode: -5$\pm$1$ ^\circ$C\\
            Electronic control box & -20 $\sim$ 40$ ^\circ$C &	6$\pm$2$ ^\circ$C\\
            Rear cover &-30 $\sim$ 10$ ^\circ$C &-30 $\sim$ -15$ ^\circ$C \\
            \noalign{\smallskip}\hline
        \end{tabular}
    \end{center}
\end{table}	

\begin{figure}
    \includegraphics[width=\textwidth, angle=0]{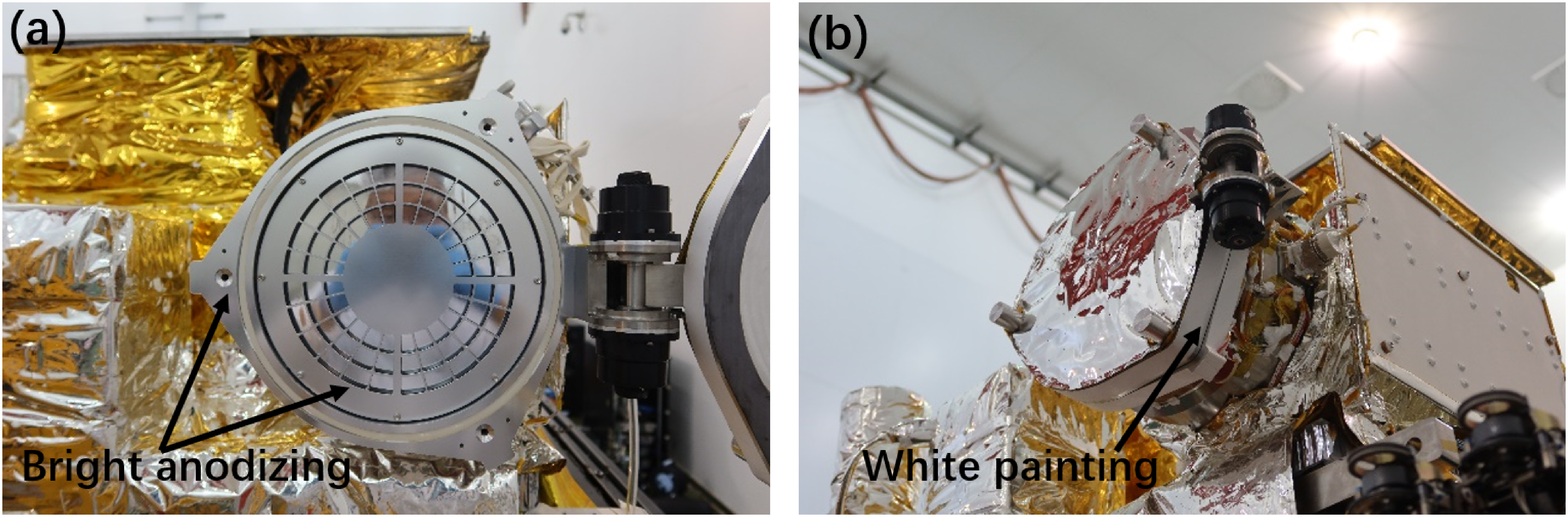} \caption{Thermal design of the front filter assembly. }
    \label{fig.9}
\end{figure}

Once the one-shot door of SUTRI is opened, the strong solar irradiation is deposited on the front thin-film filter and its supporting structure. The front filter assembly facing the Sun is treated by bright anodizing (Figure \ref{fig.9}a), with a ratio $\frac{\alpha}{\epsilon}$ of 0.38. Here $\alpha$ and $\epsilon$ represent the absorptivity and emissivity, respectively.  The bright anodizing surface reflects most of the heat flux while the rest is dissipated by white painting S781 on the side of the filter assembly seen in Figure \ref{fig.9}b. 

\begin{figure}
    \centerline{\includegraphics[width=0.5\textwidth, angle=0]{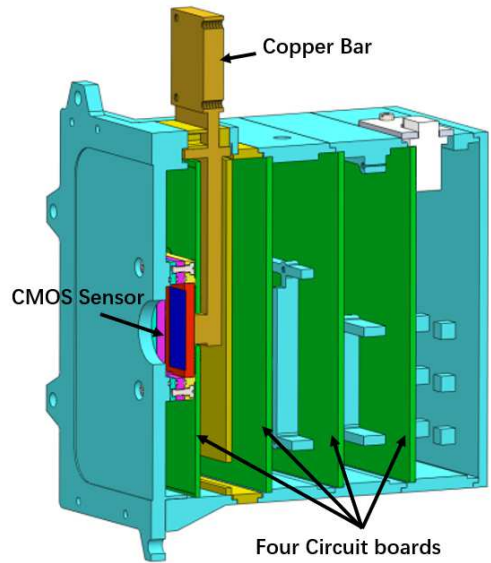}} \caption{ Mechanical design of SUTRI electric control box. Four green boards from left to right are the imaging detection, main control, primary power and inferface circuit boards, respectively. }
    \label{fig.10}
\end{figure}

Regarding the CMOS detector, we designed two working modes. One is the cleaning mode with a temperature of about 20 $ ^\circ$C to remove its absorbed contamination. The other is the normal working mode and the detector works at -5 $ ^\circ$C to reduce the dark current from the detector. The dissipation of the heat flux from the CMOS detector is conducted to the bottom radiator (Figure \ref{fig.8}a), connecting by three copper bars. The inner copper bar is from the CMOS detector \textbf{(Figure \ref{fig.10})} and the outer one connects to the outer radiator. Heaters are implemented on the outside cooper bar to realize the working temperature of the two modes. As the loaded cylinder is a sealed tube, the second cooper bar connecting inside and outside bars must be a sure-seal one. 

\subsection{Electronics and CMOS detector}
SUTRI electric control box has five types of signals with the SATech-01 satellite platform:
\textbf{
\begin{itemize}
\item Two types of output signals including the LVDS data, telemetry data (one current, two voltage and eight temperature channels);
\item Two types of input signals including the primary power supply and pulse per second signal;
\item Bidirectional signal through the RS422 communication bus.
\end{itemize}
}
From the requirement of observing modes, the motion mechanism and the monitoring LED, there are several types of internal signals generating from SUTRI’s subsystems:

\begin{itemize}
\item Output signals including focus adjustment stepper motor controlling, filter wheel step-motor controlling and LED controlling;
\item Input signals including microswitches signal from the focus adjustment Hall, filter wheel Hall signal, one-shot door Hall;
\item Bidirectional signals: CMOS imaging driving, acquisition and cache controlling signal.
\end{itemize}

\begin{figure}
    \includegraphics[width=\textwidth, angle=0]{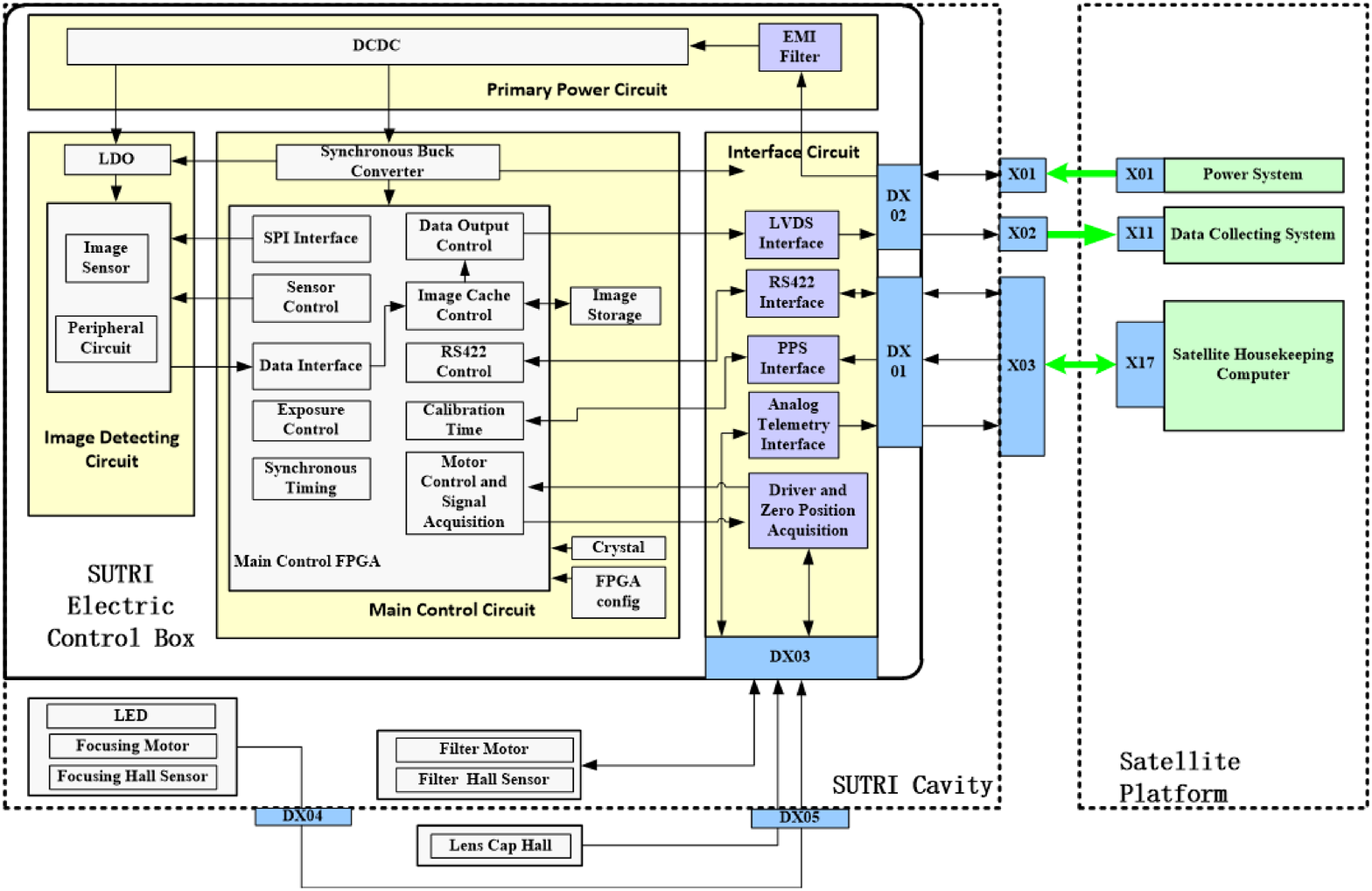} \caption{ SUTRI’s circuit functional block diagram. }
    \label{fig.11}
\end{figure}

To realize the above requirement, STURI electronic control system is composed of 5 parts, i.e., the main control, imaging detecting, primary power, interface and the front-end electronic circuits. The main control circuit has a FPGA to control SUTRI’s various subsystems and to interface with the satellite platform. The image detecting circuit is composed of CMOS sensor, CMOS power supply unit and the relative peripheral circuit. The primary power circuit includes a filter circuit, a circuit for surges suppression, a current monitoring circuit and a secondary power conversion circuit. The interface circuit includes a RS422 driver to communicate with the satellite, a pulse per second circuit, a data output driver, a step-motor driver, an acquisition circuit for position switch and a UV LED driver, etc. The front-end electronics circuit includes three micro-switches for detecting position of the one-shot door, the focus adjusting and filter wheel stepper motors. We present the circuit functional block diagram of SUTRI in \textbf{Figure \ref{fig.11}.}

The main control, imaging detecting, primary power and the interface circuits are arranged in four separate circuit boards and integrated into an electric control box, which is located at the rear side of the loaded cylinder shown in Figure \ref{fig.8}b and Figure \ref{fig.11}. In order to prevent the organic pollution, the CMOS sensor in the imaging detecting boards has been isolated from the other parts of the electronic box through indium wire. The working temperature of the CMOS sensor can also be adjusted by the thermal control system to further remove the influence of contamination, as shown in Section 3.2. The LED, the focusing adjusting and filter wheel micro-switch circuit in the front-end electronics circuit is located inside SUTRI’s loaded cylinder. The one-shot door microswitch is arranged near the door assembly. The front-end electronics circuit is connected with the electric control box through an airtight connector. 


\subsubsection{SUTRI’s Power Supply and Distribution}

The SATech-01 satellite platform power distributor provides one primary power for SUTRI with a voltage of 30 V, and an average power $\le$ 30 W. The primary power supply is converted to 5 V and $\pm$12 V through a DC-DC converter. The voltage of 5 V is further converted to the required voltage for the digital circuit by the Synchronous Buck Converters. Moreover, the voltage of 5 V is converted to the CMOS power supply voltage through LDOs. All of the power conversion circuits have over-current protection function, and the current monitoring system is designed to cooperate with satellite platform to monitor the status of the equipment.

The interface circuit of the SUTRI primary power supply is designed in strict accordance with the design and construction specifications. When the power starts, the surge current jump slope will not be higher than $10^6 \ A/s$, the surge current is not higher than 3 A, and the duration time will be less than 5 ms.

\subsubsection{SUTRI’s main control unit}
SUTRI's main control circuit consists of the main control FPGA chip, DDR2, FPGA configuration circuit and the interface circuit to monitor SUTRI’s instrumental status and operate SUTRI’s observing modes. Combining with the FPGA software, it realizes the following functions:

\begin{itemize}
\item Stepper motor control for moving the focus adjustment mechanism,
\item Microswitch for detecting the focus adjustment mechanism position,
\item Stepper motor for controlling the filter wheel,
\item Microswitch for detecting the position of five filter wheel holes,
\item Microswitch for detecting whether the one-shot door is open or closed,
\item Switching the LED on and off,
\item Imaging control of the CMOS sensor,
\item Configuration of the CMOS sensor including the gain, readout channel and exposure time parameters, etc.,
\item Arranging the imaging data from the CMOS sensor and the necessary data header parameters,
\item Sending the imaging data and the data header to SATech-01’s data collecting system with LVDS interface,
\item Receiving the telecommand from the SATech-01 satellite through RS 422 interface,
\item Arranging and sending the engineering parameters to SATech-01 satellite platform,
\item Receiving pulse per second signal and providing the on-orbit time service.
\end{itemize}

\subsubsection{SUTRI’s imaging detecting unit}

Due to the strong absorption of input irradiation, the common detector used in the visible and near-infrared cannot be used at the EUV wavelengths. Several types of sensor are employed in EUV solar detection since 1970s, such as the photographic film, frontside or backside-illuminated CCD, micro-channel plate intensifier coupled with CCD or CMOS sensors, micro-channel plate intensifier coupled with various anode readout circuits \citep{Tousey1973, Harrison1995, Delaboundini1995,Lemen2012}. Recently, CMOS imaging sensors are widely used by the EUV payloads on the latest solar missions including the SWAP onboard the Proba-2 mission \citep{Deaton2013}, EUI onboard the Solar Orbiter mission \citep{Rochus2020}. To improve the quantum efficiency of CCD or CMOS detector, the sensors need backside thinning processing. Keeping these restrictions in mind, we finally choose the GSENSE2020BSI from the Gpixel company, which is a scientific backside-illuminated CMOS image sensor with PulSar technology to detect the soft x-ray and EUV photons \citep{Harada2020}. The characteristics of GSENSE2020BSI sensor are listed in Table \ref{tab3}.

\begin{table}
    \begin{center}
        \caption[]{ Main characteristics of GSENSE2020BSI sensor.}\label{tab3}
        \begin{tabular}{clclcl}
            \noalign{\smallskip}\hline
            Parameter & Value\\
            \noalign{\smallskip}\hline
            Pixel size & 6.5 microns\\
            Number of pixels &2048 $\times$ 2048 \\
            ADC	& 12 bit \\
            Frame rate &	3 frames/second with the exposure time less than 1 second \\
            Gain &	High Gain 1: 0.47 e-/DN \\
            & High Gain 2: 5.29 e-/DN \\
            & Low Gain 3: 11.62 e-/DN \\
            &Low Gain 4: 17.42 e-/DN \\
            Full well capacity & 	High Gain 1:  1.8 ke- \\
            & High Gain 2:  17 ke-\\
            & Low Gain 3:  34 ke-\\
            & Low Gain 4:  57 ke-\\
            Read out noise& High Gain 1:  4 e-\\
            & High Gain 2:  23 e- \\
            & Low Gain 3:  35.9 e- \\
            & Low Gain 4:  64 e- \\
            Quantum efficiency& $\ge20\% @$46.5nm \\
            
            \noalign{\smallskip}\hline
        \end{tabular}
    \end{center}
\end{table}	

\begin{figure}
    \includegraphics[width=\textwidth, angle=0]{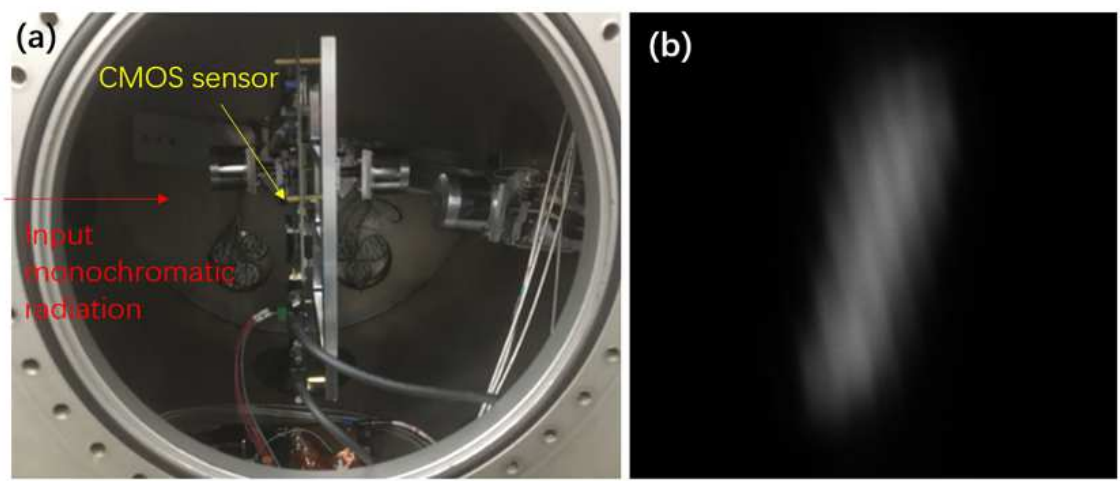} \caption{ Panel (a): Photos of the experimental system in the calibration chamber of the National Synchrotron Radiation beamline used to measure the quantum efficiency of GSENSE2020BSI CMOS sensor. Panel (b): The monochromatic light recorded by GSENSE2020BSI sensor passing through the monochromator exit slit at 46.5 nm.  }
    \label{fig.12}
\end{figure}

\begin{figure}
    \includegraphics[width=\textwidth, angle=0]{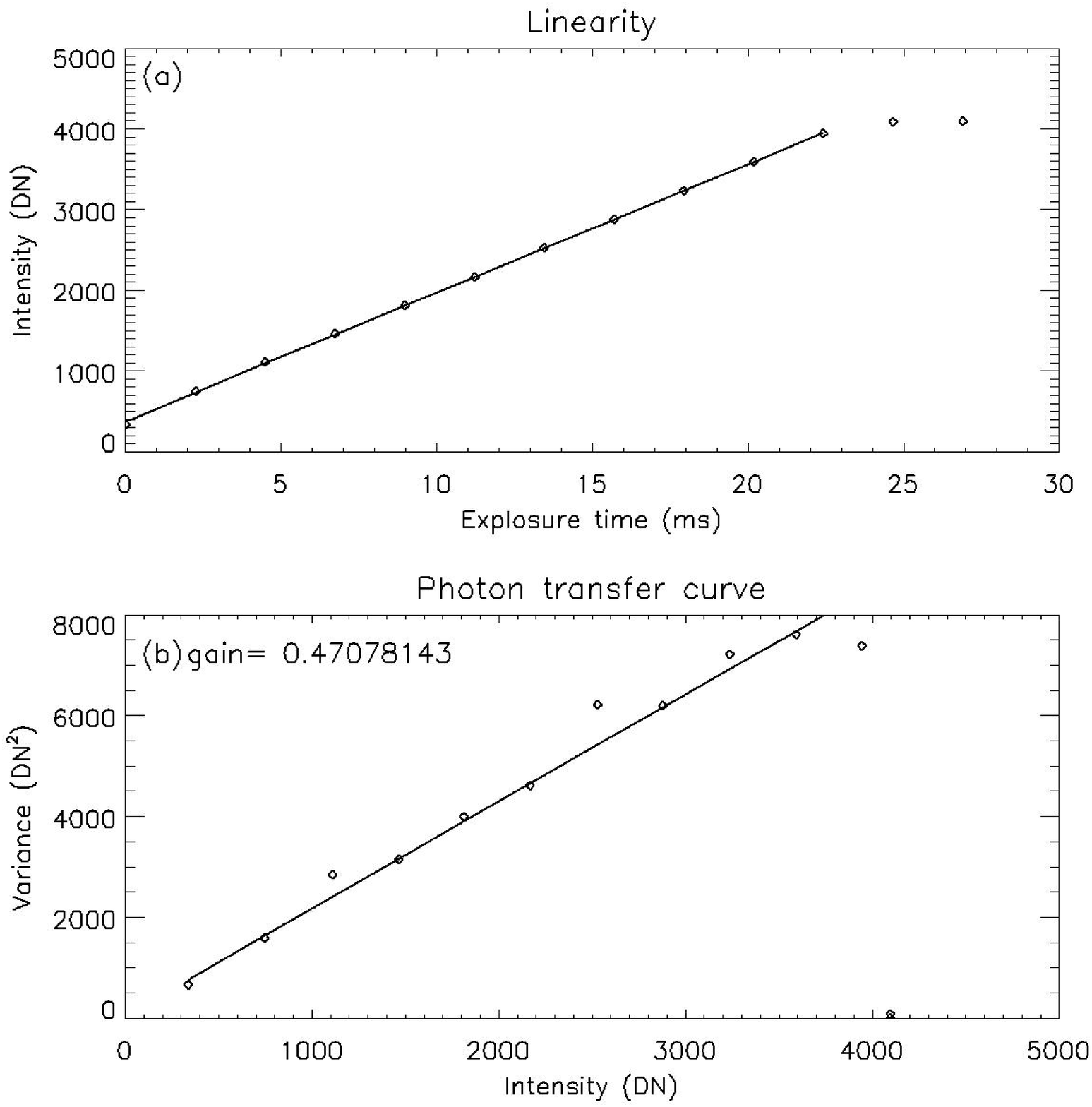} \caption{The linearity and the photon transfer curve (diamond symbols) of SUTRI’s CMOS camera for the high gain 1 channel after the subsystem integration with uniformed visible light source produced by an integrating sphere. The gain value in panel b is derived from the slope of a linear fitting (solid line) to the diamond symbols.  }
    \label{fig.13}
\end{figure}

Since China has very limited experience in EUV astronomical observations and the GSENSE2020BSI sensor has never been used by any Chinese ground EUV instruments or spaced-based EUV payloads, extensive tests have been made at the sensor level. In 2020, we firstly tested its performance with an evaluation electronic boards provided by Gpixel company in the laboratory, focusing on its performance on the gain, readout noise and the full well capacity. These tests were done with the integrating sphere working at visible wavelength. Once the parameters have been determined, we set up an experimental system to make sure that the GSENSE2020BSI works well in the vacuum and is able to transfer imaging data to the computer working in the air through vacuum feedthroughs. We took the system to soft x-ray and UV beamline of the Chinese National Synchrotron Radiation Laboratory, operated by the University of Science and Technology of China (Figure \ref{fig.12}a). From the monochromatic synchrotron radiation light passing through the monochromator's exit slit recorded by our CMOS sensor(Figure \ref{fig.12}b), we believe that GSENSE2020BSI senor is able to detect EUV photons at 46.5 nm. We also calibrated the quantum efficiency of GSENSE2020BSI sensor, and the primary result indicated that its value is larger than 20$\%$ at 46.5 nm. Due to the contamination on the GSENSE2020BSI detector and the calibrating detector, as well as the high order radiation from the grating momochromator at the wavelengths lower than 46.5 nm, the systematic error for the quantum efficiency measurement should be large. We cannot obtain an accurate result by now. 

We then started to develop SUTRI’s CMOS camera after the evaluation of GSENSE2020BSI sensor, which is installed on the imaging detecting circuit board in SUTRI's electronic box. The imaging detecting circuit board provides the power supply of the GSENSE2020BSI CMOS sensor and its configuration is set by the FPGA on the main control board. The power of the sensor is about 1.2 W. As shown in Section 3.3, the heat flux from the sensor conducted to the backside irradiator by three copper bars. Our CMOS sensor has two individual data output channels with high and low gains. Each gain channel has 4 pairs of LVDS output channels with a data rate of 300 Mbps. After DDR2 data storing, the data from the sensor is sent to the satellite platform through LVDS with a data rate of 50 Mbps. The data from the sensor is 12 bit and is written to 16 bit, hence the data size of one image is 8 MB. One image is transferred to the satellite within 1.5 s. The frame rate of SUTRI camera is about 3 frames per second if the exposure time is less than 1 s. 

We firstly tested the performance of the CMOS camera after the electronic box was integrated at sub-system level. The measured gain, readout noise and full well capacity values can be found in Table 3. The gain is derived from photon transfer curve (Figure \ref{fig.13}b) and the linearity (Figure \ref{fig.13}a) is further obtained with the same data \citep{Duan2020}. Four different gain configurations were determined to adapt various solar transition region features. For example, the gain with 0.47 e-/DN  is suitable for observations of quiet network regions because its readout noise is very low. The gain with 17.42 e-/DN has the largest dynamical range and can be used to observe active regions and solar flares. The performance of the CMOS was monitored at the SUTRI’s system integration and test phase. The dark current during the thermal vacuum test was measured and it is upgraded every day during SUTRI’s on-orbit observation.

\section{Operation and on-orbit Calibration }

\subsection{Operation}
\begin{figure}
    \includegraphics[width=\textwidth, angle=0]{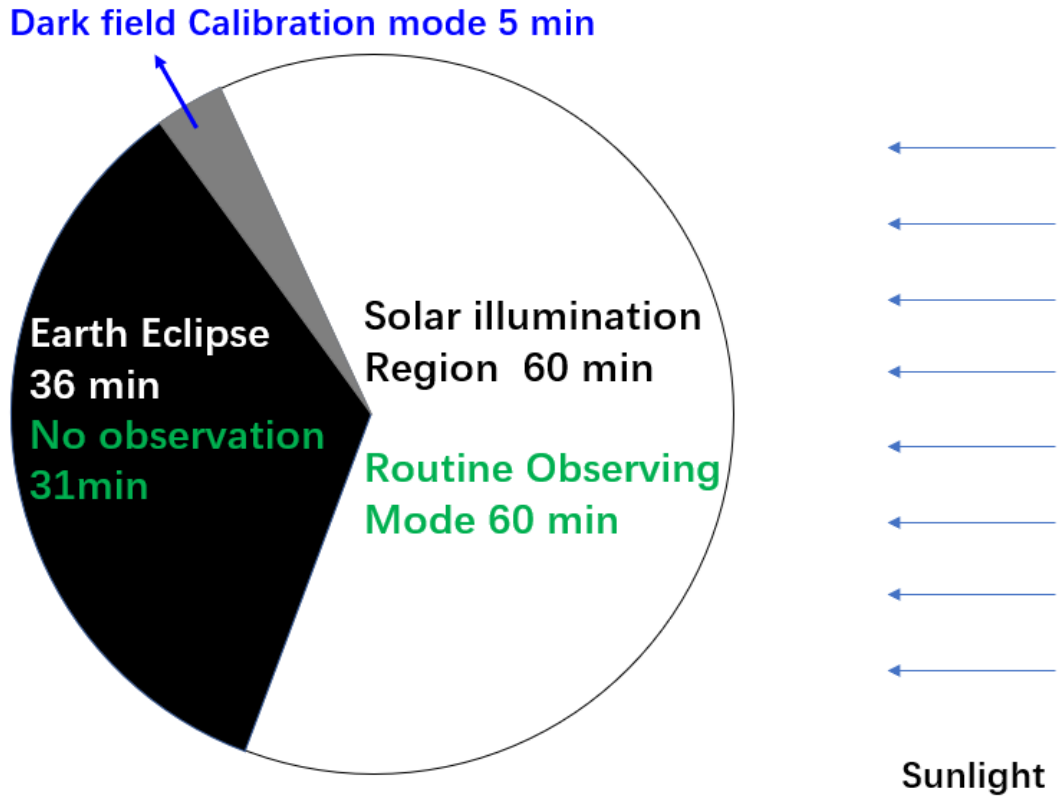} \caption{Routine observation strategy of SUTRI for a 96 min orbit period. }
    \label{fig.14}
\end{figure}

The SATech-01 satellite is not a solar dedicated mission and its orbit is not optimized for solar observations. Normally, the orbit period is about 96 min. Each orbit is divided into two parts, i.e., the solar illumination and earth eclipse regions as the local time at the descending node of the orbit is 10:30 am. During solar illumination time lasting for 60 min, SATech-01 points to the Sun and SUTRI takes routine solar observations or flat field calibration observations. The satellite is in the earth eclipse for the rest 36 min (Figure \ref{fig.14}) and SUTRI takes dark field calibration observations for about 5 min. SUTRI does not take images for the remaining 31 min. If there are data download windows (once per day) or the other payloads on the SATech-01 satellite do not need solar pointing in solar illumination time, the satellite does not point to the Sun but SUTRI still takes observations. The obtained data at this moment are also used for generating dark fields. 

SUTRI’s observation is realized by the observing plans containing a list of spacecraft and instrument telecommands and their execution time is generated by the ground operation team using the timeline tool. There are four telemetry windows each day that we are able to upload the observing plans. Generally, one window is used to upload the SUTRI’s telecommands because the observing plan is updated every day. The other windows are used for emergency or special scientific observations. The full-disk solar imaging data is downloaded twice every day from the satellite to the ground without compression. 

Table 1 shows that SUTRI has four observing modes, i.e., the routine observing mode, the dark calibration mode, the flat field calibration mode and the cleaning mode. SUTRI’s observing plans are the combination of these four modes. For the routine observing mode, the normal configurations include the following functions:

\begin{itemize}
\item Configure the thermal control system to work at normal working mode and wait for thermal balance,
\item Point the satellite to the Sun,
\item Power on SUTRI and finish the initialization process,
\item Choose the suitable filter wheel (Al/Mg/Al filter with the thickness of 77/102/77 nm is used for routine observation) and move the focus stepper motor to the position with best focus,
\item Configure the exposure time, gain parameters and the read out channel,
\item Take solar images with a cadence repeatedly for each orbit including 60 min during the solar illumination part and 5 min during the earth eclipse time.
\end{itemize}
As a default setting SUTRI is taking full-disk images at a cadence of 30 s and the exposure time can vary from 1 to 25 s with the normal cadence. A long exposure up to 300 s is also supported. If needed, we can adjust the exposure time and cadence for coordinated observations with other telescopes. SUTRI does not have an automatic exposure control, so we need to configure the suitable gain and exposure time for solar flares observations to avoid saturation.

\subsection{On-orbit Calibration}
The best focus of SUTRI can be determined on-orbit by taking a series of solar images at different focus positions. As the on-orbit spatial resolution of SUTRI is almost the same as ground testing and the pointing of the satellite slightly changes during the exposure for single image, the focus position has not been changed so for. 
 
Since the CMOS sensor works at a low temperature less than 0$ ^\circ$C for normal thermal controlling mode, it is easy to deposit contamination, such as the residual water vapour. The cleaning mode is designed in the thermal control system. Once there is obvious degradation from the CMOS sensor, the SUTRI  will work at cleaning mode for a period of time. It returns back to the normal thermal controlling mode and we can start normal solar observations then.  
  
The bias and dark current of the CMOS sensor changes with time due to the on-orbit degradation of the senor. Moreover, the pixel-to-pixel response of the CMOS sensor is different because each pixel has its own amplifier. So we design the dark calibration and the flat field calibration modes to calibrate the dark current and the instrumental pixel-to-pixel nonuniformity. The dark field is taken during earth eclipses. The flat-filed calibration is realized by pointing the satellite  to 8 off-pointing positions slight away from normal solar pointing, similar to that generally used by TRACE, AIA \citep{Lemen2012, Handy1999} and the other full-disk solar telescopes \citep{Li2019,Su2019}.

\begin{figure}
    \includegraphics[width=\textwidth, angle=0]{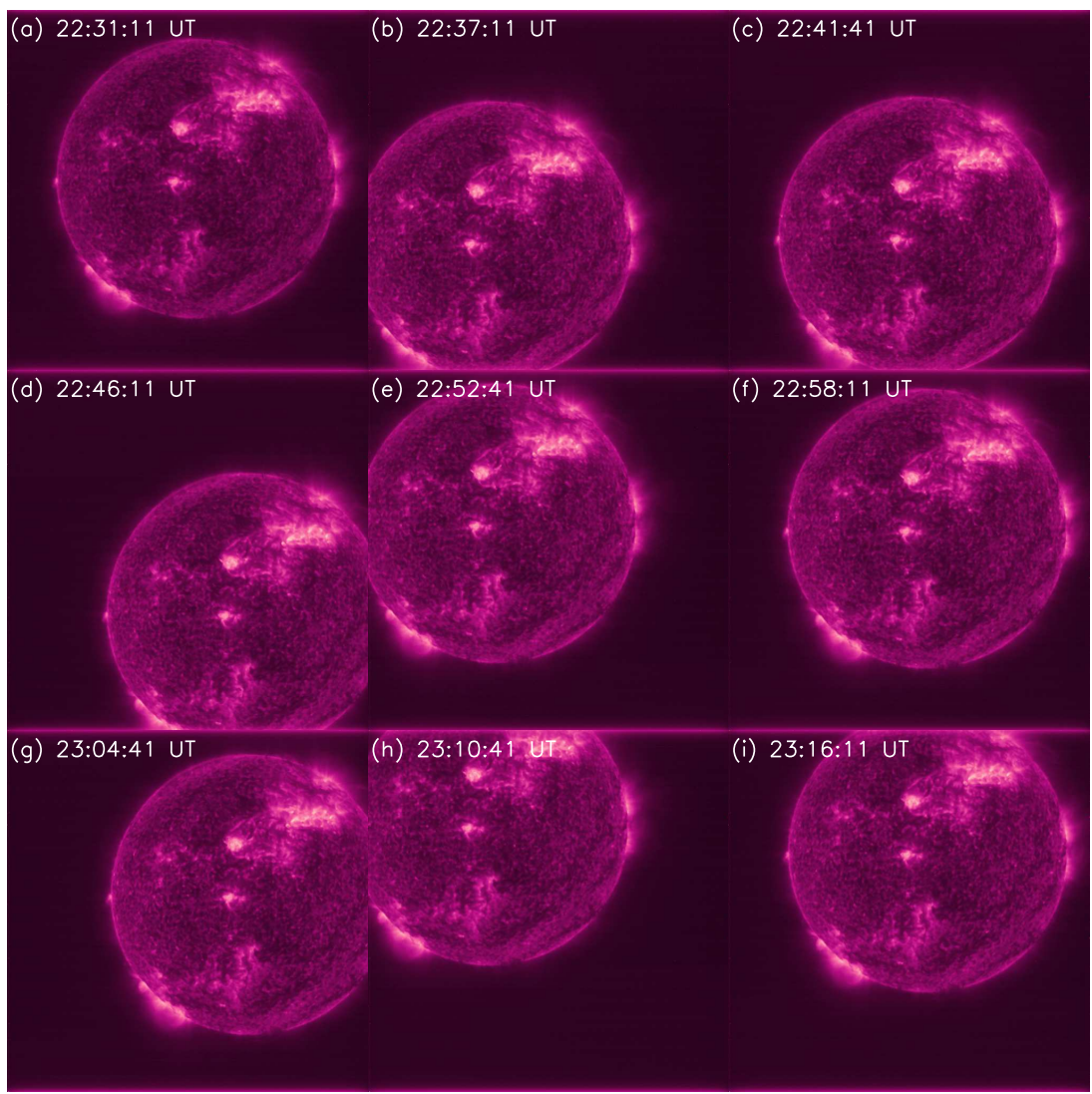} \caption{One normal (panel a) and eight off-pointing images taken during the flat-field calibration mode. }
    \label{fig.15}
\end{figure}

\section{Data processing and data availability}
All of the recorded data from the SATech-01 satellite is firstly received twice per day by the satellite operation team from the Innovation Academy for Microsatellites, Chinese Academy of Sciences. SUTRI’s imaging and telemetry data is further parsed, renamed with SUTRI’s file naming rule and transferred to the SUTRI science operation and data analysis team from NAOC and Peking University within one or two days. High-level calibration and scientific data are produced with our data processing pipeline by the data server at Huairou Solar Observing Station (HSOS) of NAOC. The quick-look and high-level science data is stored in the SUTRI archive, released to the public free of charge and can be download through the SUTRI website (https://sun10.bao.ac.cn/SUTRI/).

\subsection{Data Levels}
SUTRI data is classified into Level 0, Level 0.9, Level 1 and Level 1.5. The data with level larger than level 0.9 will be released. The original binary imaging and telemetry data of SUTRI downloaded from the satellite is produced by the satellite operation team in bin and xlsx formats, respectively. Once the binary imaging data and the telemetry data are received by the HSOS data server, we generate these data to Level 0.0 fits file, which has 2kx2k size with 16 bit and the fits header is also generated containing the information about exposure time, gain value, and filter position et al.. Level 0 data is also classified to calibration data and scientific data according to the observing plan and modes and stored in the corresponding directories. The time in Level 0 fits file is transformed to UTC time, which will be further used by the higher-level data.
 
Level 0.9 and 1.0 data are the calibrated routine observing fits files that can be used for scientific analysis. The pixel value in Level 0.9 and 1.0 has DN unit. The data users can transform the pixel value in DN unit to electrons according to the gain value in the fits header if they want to know the electrons. Combining with the exposure time, the electron numbers per pixel per second for a solar feature can be obtained. The data level higher than 0.9 uses the helioprojective-cartesian coordinate, rather than the detector coordinate used in Level 0. 

Level 1.5 data will be calibrated to physical units after the radiometric calibration. Level 0.9, 1.0 and 1.5 files can also be generated to MPEG movies, which are the quick look data. Data users are recommended to downloaded these quicklook movies from SUTRI data archive. If certain interesting solar events are found, they can download the corresponding fits files for further analysis. The first released SUTRI data will be level 0.9.  Below we summarize the data processing flow briefly.

\subsection{Data processing flow}
SUTRI’s data processing includes the process of routine and calibration mode data. Here we focus on the data production from Level 0.0 to 0.9. As SUTRI does not have an imaging stabilization system and the SATech-01 pointing accuracy is 0.005$^\circ$, we made great efforts to deal with the jitters and rotation among SUTRI’s image sequence.

Calibration mode data processing including the generation of dark and flat field calibration file. The data observed during the earth eclipse is firstly identified with the value less than a threshold value according to the gain configuration. These data are stacked together and the mean value is calculated for each pixel to generate dark fields. The exposure time of the dark field is the same as that for the solar observation, so the dark fields contains the dark current, bias values from the detector and the read noise. For the processing of flat field calibration file, the off-pointing images are firstly corrected for dark fields. The dark corrected off-pointing images are further aligned to the first image and the offset values between different off-pointings are determined. Then we use the algorithm proposed by \cite{Chae2004} and \cite{Kuhn1991} to generate the flat-field, which is widely used by many ground and space-based full-disk solar telescopes \citep{Li2022,Li2021,Lemen2012,Su2019}.

Data processing procedure of routine data from Level 0.0 to Level 0.9 is as follows:

\begin{itemize}
\item Correct the dark field, which is updated every day. As the CMOS sensor works at about -5 $ ^\circ$C, the dark current is negligible and the dark field is dominated by the bias value.

\item Remove the horizontal stripes from the row-by-row CMOS readout circuit. SUTRI’s CMOS camera is readout row by row after exposure and horizontal stripes with the values ranging from 1 to 3 DN relative to the mean value of the whole image can be found in the original images. The horizontal stripes are removed with the wavelet analysis method.  

\item Identify and fix the hot, bad pixels and the spikes influenced by cosmic ray and high energy particles on the CMOS detector. Bad and hot pixels are found from both the ground and on-orbit images and the number of hot pixels depends on the working temperature of CMOS sensor. As SATech-01 satellite operates at a sun-synchronous orbit, the satellite passes through the south and north poles of the earth every orbit and also the South Atlantic Anomaly occasionally. The hot pixels and spikes influenced by high energy particles increase a lot in these regions. As the values of hot, bad pixels and the spikes are statistically different from those of the normal pixels, we identify these pixels automatically and replace them with the median values of neighboring pixels, similar to the algorithm used for AIA \citep{Lemen2012}. 

\item Identify and fix the CMOS pixels occluded by dust. Dark dots or patches are found in SUTRI images possibly caused by dust during the ground integration and test phase. The on-orbit number of dots also changes with time, possibly relate to the solid lubrication film coated on the motion mechanism. The values in the dark spots are almost the same as dark field and cannot be efficiently corrected by the flat-fielding processing. We firstly identify the positions of these dark dots and set up a template to mark the positions. A low-pass filter is employed for SUTRI’s images after the above-mentioned processing. The values of the dark dots are replaced by the corresponding intensities in the low-pass filtered image.
  
\item Align the series of solar images and correct for the imaging shift and rotation caused by the variation of the satellite pointing. The star tracker of SATech-01 satellite does not work during parts of the orbit when there are not enough reference stars. The pointing and stability of the satellite become worse at the time and one can see the drift of the solar disk in the original images and even part of solar disk is invisible (no more than ten min per orbit). The data quality is evaluted and the images with bad qualities such as blurred images are dropped. The disk center is further determined by fitting the solar limbs and the drift and rotation between series of SUTRI’s images are derived with optical flow method.  
  
\item Convert the coordinate from the detector coordinate system to the helioprojective-cartesian coordinate system. With the determined solar disk center positon, the disk center of every image is shifted to the center of the 2kx2k image array. The solar north is flipped to the top of the array including the correction of P angle. As the roll angle of the satellite is not used, the solar north has a slight difference from that the data from other solar mission, 

\item Standardize of filename and header information.
 
\end{itemize}

\begin{figure}
    \includegraphics[width=\textwidth, angle=0]{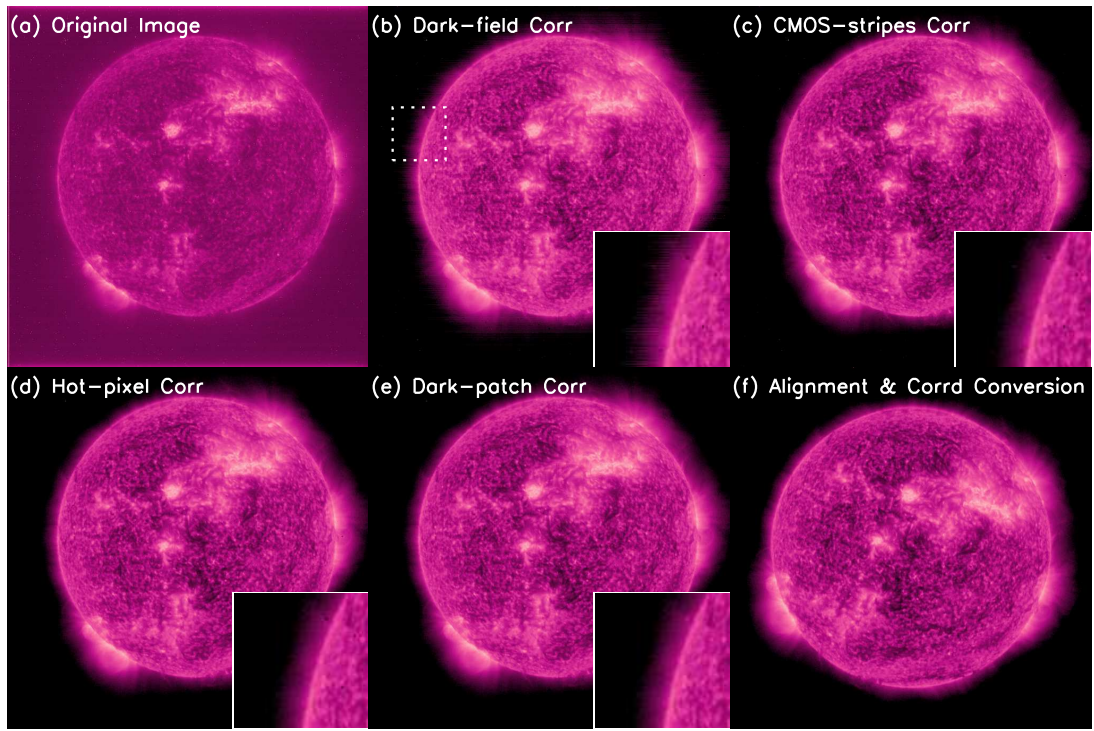} \caption{An example of SUTRI's Data processing from Level 0 to Level 0.9. To better illustrate the improvement of the scientific data in each step, the sub-images at the right and bottom corner of Panels b, c, d and e show the zoom-in images of the white box region marked in panel b.  }
    \label{fig.16}
\end{figure}

Figure \ref{fig.16} shows an example of the data processing from Level 0 to Level 0.9. We have not done the flat-field correction in the current data processing pipeline. Once the flat-field correction is added to the data processing pipeline, the data level will be updated to Level 1.0.

\section{Conclusion}
The SUTRI door was opened on Aug. 30, 2022. The first solar transition region image at 46.5 nm was obtained on Aug. 31. We then adjusted the exposure time and took the solar images with different rear filters for test. SUTRI has performed routine observations of the Sun since Sep. 4, 2022 with a spatial resolution of $\sim$8" and a cadence of 30 s. Various types of solar activities such as coronal jets, flares, filament and prominence eruptions are captured. Prevalent downflows with a temperature of 0.5 MK are found at the legs of active region coronal loops \textbf{\citep{wu2023}}. We also see the network features in transition region and coronal holes in SUTRI’s images. After four months on orbit, SUTRI works well now and more than 1.6 TB data (about 200, 000 frames) has been recorded so far. The newly developed Sc/Si multi-layer reflecting mirror and the CMOS camera are validated on-orbit. We are also taking routine calibration data to monitor long-term behaviors of SUTRI. The data processing pipeline is developed and high-level scientific data is generated and released to the public on Jan. 11, 2023 through the Internet.

\begin{acknowledgements}
SUTRI is a collaborative project conducted by the National Astronomical Observatories of CAS, Peking University, Tongji University, Xi’an Institute of Optics and Precision Mechanics of CAS and the Innovation Academy for Microsatellites of CAS. The authors are supported by NSFC grants 11825301, 12003016 and the Strategic Priority Research Program of the Chinese Academy of Sciences with the Grant No. XDA15018400.
\end{acknowledgements}




\bibliographystyle{raa}
\bibliography{ms}
\label{lastpage}

\end{document}